\documentclass[JKPS,twoside,twocolumn,epsfig ]{revtex4}
\usepackage{graphicx}

\begin{document}
\title[] {%Probit and logit models in Korean financial markets
Dynamical Structures of High-Frequency Financial Data }
\author{Kyungsik \surname{Kim$^{1}$}}
%\thanks{Electric mail: kskim@pknu.ac.kr;\\
%Tel: +82-51-620-6354; Fax: +82-51-611-6357}
\author{Seong-Min \surname{Yoon$^{2}$}}
\author{SooYong \surname{Kim$^{3}$}}
\author{Ki-Ho \surname{Chang}$^{4}$}
\author{Yup \surname{Kim$^{5}$}}
\affiliation{
$^{1}$Department of Physics, Pukyong National University,\\
Pusan 608-737, Korea \\
$^{2}$Division of Economics, Pukyong National University,\\
Pusan 608-737, Korea\\
$^{3}$Department of Physics, Korea Advanced Institute\\
of Science and Technology, Daejeon 305-701, Korea\\
$^{4}$Remote Sensing Research Laboratory, Meteorological Research
Institute, KMA, Seoul 156-720, Korea\\
$^{5}$Department of Physics, Kyung Hee University,\\
Seoul 130-701, Korea \\
}

%\received{ February 2005 }

\begin{abstract}

We study the dynamical behavior of high-frequency data from the
Korean Stock Price Index (KOSPI) using the movement of returns in
Korean financial markets. The dynamical behavior for a binarized
series of our models is not completely random. The conditional
probability is numerically estimated from a return series of KOSPI
tick data. Non-trivial probability structures can be constituted
from binary time series of autoregressive (AR), logit, and probit
models, for which the Akaike Information Criterion shows a minimum
value at the $15$th order. From our results, we find that the value
of the correct match ratio for the AR model is slightly larger than
the findings of other models.
\\
\hfill\\
PACS numbers: 89.65.Gh, 05.40.-a, 05.45.Df, 89.65.-s\\
%$Keywords$: Correlation function, Conditional probability
%distribution, AR model, logit and probit models, KOSPI

\end{abstract}

\maketitle

Recent investigation of differently scaled economic systems has been
received a considerable attention as an interdisciplinary field of
physicists and economists $[1,2,3,4,5,6,7,8]$. One of challenging
issues is to test efficient market hypotheses from the perspective
of empirical observations and theoretical considerations. To exploit
or predict the dynamical behavior of continuous tick data for
various financial assets $[9,10]$ is extremely desirable. Financial
efficiency and predictability can significantly benefit investors or
agents in the financial market and successfully reinforce the
effective network between them. For example, when the price of stock
rises or falls in the stock market, a trader's decision to buy or
sell is influenced by various strategies, external information, and
other traders. One such strategy is to apply the up and down
movement of returns to a correlation function and the conditional
probability. This strategy, which is pivotal for predicting an
investment, is a useful tool for understanding the stock
transactions of company whose stock price is rising or falling. In
the literature, Ohira et al. $[9]$ mainly discussed conditional
probability and the correct match ratio of high-frequency data for
the yen-dollar exchange rate; they showed that such dynamics is not
completely random and that a probabilistic structure exists. Sazuka
et al. $[10]$ used the order $k=10$ of the Akaike Information
Criterion (IC) to determine the predictable value of the
autoregressive (AR) model; in contrast, they numerically calculated
the $5$th order of the logit model $[11]$. Motivated by such
research, we apply and analyze novelly the AR, logit, and probit
models to the Korean financial market, which, in contrast to active
and well-established financial markets, is now in a slightly
unstable and risky state.

Interest in nonlinear models has recently grown, particularly in the
social, natural, medical, and engineering sciences. Statistical and
mathematical physics provides a powerful and rigorous tool for
analyzing social data. Moreover, several papers have focused on
social phenomena models based on aspects of stochastic analysis,
such as the diffusion, master, Langevin, and Fokker-Planck
equations. Many researchers in econometrics or biometrics have
proposed the use of AR, logit, and probit models in the formulation
of the discrete choices, including binary analysis. Interestingly,
Nakayama and Nakamura $[16]$ associated the fashion phenomena of the
bandwagon and snob effects with the logit model. To our knowledge,
in addition to the Akaike IC, there are at least two other similar
standards such as the Hannan-Quinn IC and the Schwarz IC. However,
we restrict ourselves to find the Akaike IC as the residual test in
order to minimize the remained value for binary analysis. Moreover,
after calculating the binary structures and their Akaike IC value,
we compute the correct match ratio, or the power of predictability.
Although the dynamical behavior of logit and probit models has been
calculated and analyzed in scientific fields such as mathematics,
economics, and econophysics, until now these models have not been
studied in detail with respect to financial markets

In this letter, we present the future predictability function of the
AR, logit, and probit models, by using the tick data analysis of the
Korean Stock Price Index (KOSPI) for the Korean financial market. By
examining the binary phenomena of a financial time series in terms
of the nontrivial probability distribution, we show that the
high-frequency data of our model follows a special conditional
probability structure for the up and down movement of returns.
Moreover, our results are of great importance for making a powerful
and capable tool that can be used to investigate properties of
efficient and predictable markets.

In our calculations, the return of the tick data at time $t$ is
$R(t)= \ln p(t + \triangle t)/ p(t )$ for the price $p(t)$, and the
return change is $ D(t)\equiv R(t+1)-R(t)$ for every time $t$. From
the series of tick data in one asset, we can binarize the $\{X(t)\}$
series as follows: $X(t)=+1$ if $D(t)>0$ and $X(t)=-1$ if $D(t)<0$.
We can then extend the $\{X(t)\}$ series to a random walk formalism
as $Z(t+1)=Z(t)+X(t)$. Moreover, we can determine the cumulative
probability distribution and the conditional probabilities from the
random walk of the one-directional zigzag motion.
The correlation function can also be calculated as
\begin{equation}
C(u)=<D(t+u)D(t)>. \label{eq:d4}
\end{equation}
\begin{figure}[]
\includegraphics[angle=90,width=8.5cm]{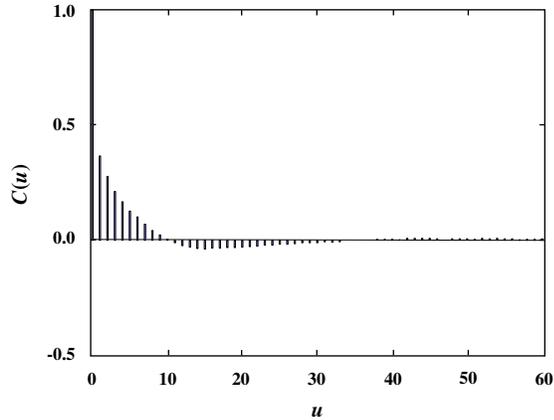}
\caption[0]{Plot of the correlation function, $C(u)$, from the set
of minutely tick data, Data $A$, of the KOSPI; the data were
collected from January $1997$ to December $1998$.}
\end{figure}

We now introduce the AR, logit, and probit models $[11,12,13,14]$
for an $\{X(t)\}$ series of continuous tick data. The AR model is
defined by
\begin{equation}
\textrm{AR}(k)=\alpha_{0} + \sum_{i=1}^{k} \alpha_{i} X(t-i) +
\epsilon (t), \label{eq:d44}
\end{equation}
where $\epsilon (t)$ is a white noise with Gaussian distribution
of zero mean and variance $\sigma$. The standard logit model for
binary analysis $[12]$ is described as
\begin{equation}
\log\textrm{it}(p) = \log \frac{p}{1-p} = \beta_{0} + \sum_{i=1}^{k}
\beta_{i} X(t-i) + \epsilon (t), \label{eq:e55}
\end{equation}
where $p$ is a dummy variable between $0$ and $1$. The linear probit
model from Eq. $(3)$ is represented in terms of
\begin{equation}
\textrm{probit}(p)={\Phi}^{-1} (p) = z_p \label{eq:g77}
\end{equation}
where ${\Phi}^{-1} (\cdot)$ is the inverse of the standard normal
cumulative distribution function, and the standard normal cumulative
distribution function is given by $ {\Phi}(z_p ) = \textrm{Pr}(z\leq
z_p )= (1/\sqrt{2\pi}) \int^{z_p}_{-\infty} dz \exp({-z^{2}} /2)$.
Furthermore, we make use of Eqs. $(2)-(4)$ to find out binary
structure and its correct match ratio, and these mathematical
techniques lead us to more general results of predictability. To
determine the minimized order $k$ of our model, we define the Akaike
IC $[12,13]$ as
\begin{equation}
\textrm{AIC}=\frac{2}{T} [-\ln {Ml} + \ln {Mp}] \label{eq:g9}
\end{equation}
for the sample size $T$, where $Ml$ and $Mp$ stand for the maximum
likelihood and the number of parameters, respectively.

%\section {Numerical results }

To analyze the correlation function and the conditional probability,
we introduce our underlying asset into the KOSPI in the Korean
financial market. First, we consider two delivery periods: the first
set of data, Data $A$, was from January $1997$ to December $1998$;
the second set, Data $B$, was from January $2004$ to December
$2004$. The lag time of two sets of tick data is about one minute.
Data $A$ contains $133,823$ items of data and Data $B$ contains
$86,561$ items.

\begin{table}[]
\caption[0]{Values of conditional probability from the simulation
results of  Data $A$ and Data $B$; NP stands for the number of tick
data points.}
\begin{tabular}{lcr}\hline
KOSPI & Data $A$ & Data $B$   \\ \hline
 NP            &$133,823$  & $86,561$ \\ \hline
$P(+)$          &$48.65$  &$49.80$    \\
$P(+|+)$        &$57.85$  &$50.37$    \\
$P(+|-)$        &$39.93$  &$49.23$    \\
$P(+|+,+)$      &$68.48$  &$52.25$    \\
$P(+|+,-)$      &$53.10$  &$51.62$    \\
$P(+|-,+)$      &$43.26$  &$48.46$    \\
$P(+|-,-)$      &$31.18$  &$46.92$    \\
$P(+|+,+,+)$    &$74.70$  &$53.27$    \\
$P(+|+,-,+)$    &$50.90$  &$48.98$    \\
$P(+|+,-,-)$    &$42.98$  &$48.41$    \\
$P(+|-,-,+)$    &$34.61$  &$47.90$    \\
$P(+|-,-,-)$    &$25.83$  &$45.61$    \\
$P(+|+,+,+,+)$  &$77.51$  &$54.04$    \\
$P(+|+,+,+,-)$  &$66.42$  &$53.05$    \\
$P(+|+,+,-,+)$  &$56.17$  &$50.10$    \\
$P(+|+,-,-,+)$  &$40.60$  &$47.48$    \\
$P(+|+,-,-,-)$  &$34.24$  &$46.12$    \\
$P(+|-,+,+,+)$  &$66.39$  &$52.39$    \\
$P(+|-,-,+,-)$  &$40.21$  &$49.88$    \\
$P(+|-,-,-,+)$  &$30.09$  &$48.31$    \\
$P(+|-,-,-,-)$  &$22.90$  &$45.19$    \\
\hline
\end{tabular}
\end{table}

From the two tick data, we computed two series: the X(t) series and
the Z(t) series, where Z(t) represents a one-dimensional zigzag
motion. This computation refers to a binary strategy of the buy and
sell trend of traders in financial markets. Fig. $1$ plots the
correlation function ,$C(u)$, which we obtained from the return
change $ D(t)$. The plot suggests that the minutely returns for Data
$A$ of the KOSPI are not entirely independent of, or different from,
the random walk model but almost independent for long periods. Given
the probabilistic structure of our model, we can deduce from the
correlation function that the dynamical behavior is completely
nonrandom.

\begin{figure}[]
\includegraphics[angle=90,width=8.5cm]{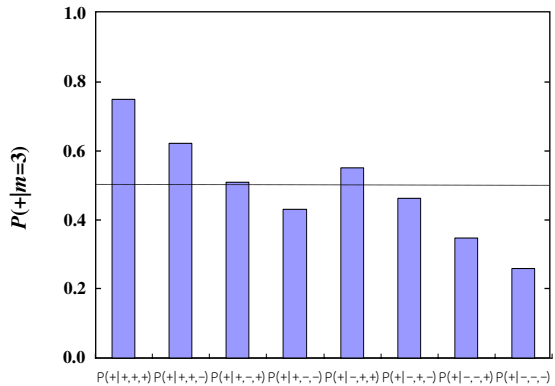}
\caption[0]{Conditional probabilities $P(+|m=3)$ for the set of
minutely tick data, Data $A$, of the KOSPI.}
\end{figure}

\begin{figure}[]
\includegraphics[angle=90,width=8.5cm]{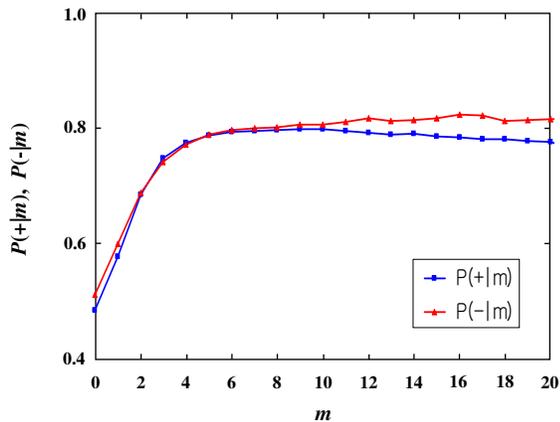}
\caption[0]{Plot of conditional probabilities $P(+|m)$ and $P(-|m)$
for the set of minutely tick data, Data $A$, of the KOSPI.}
\end{figure}

By quantitative analysis, we can relate the $X(t)$ series to
conditional probability. To analyze the high-frequency data of the
KOSPI, we concentrated on the up and down return movements in terms
of conditional probability. The parameter $P(+|+,+)$ refers to the
conditional probability that a tend in the price returns is likely
to move in the same direction; that is, that the price is likely to
rise after two consecutive steps in the same direction. Table $1$
summarizes the results of various conditional probabilities for Data
$A$ and Data $B$ of the KOSPI. Fig. $2$ shows that the conditional
probability of $P(+|+,+,+)$ has a remarkably larger value than the
probability of $P(+|m=3)$, except $P(+|+,+,+)$. From our results, we
can give the relation of the three parameters as $P(+|+,+)=p$,
$P(+|-,+)=q$, and $P(+|+,+,+)=p+\alpha$ for $0<\alpha< p <1$,
$0<q<1$. Figure $3$ shows that the conditional probability $P(+|m)$
($P(-|m)$) has a larger value than $P(+|m-1)$ ($P(-|m-1)$), which
exists for one selling state or buying state after $m-1$ selling
states or $m-1$ buying states. When we compare this result to that
of the yen-dollar exchange rate of the Japaneses financial market,
our conditional probabilities for $m<5$ have a slightly larger value
than those of the yen-dollar exchange rate $[9]$. The values of
$P(+|m)$ and $P(-|m)$ for $m<6$ increase continuously while the two
values for $m\geq 6$ are almost constant; in this case, the period
of the $m$ states is about $m$ minutes in real times. We predict
this result to be consistent with the buy-sell strategy of dealers
who can change in a few minutes. Note that although Data $A$ and
Data $B$ share a significant similarity, we cannot understand the
behavior of these data sets from a random walk model that has fixed
values for conditional probabilities.

\begin{figure}[]
\includegraphics[angle=90,width=8.5cm]{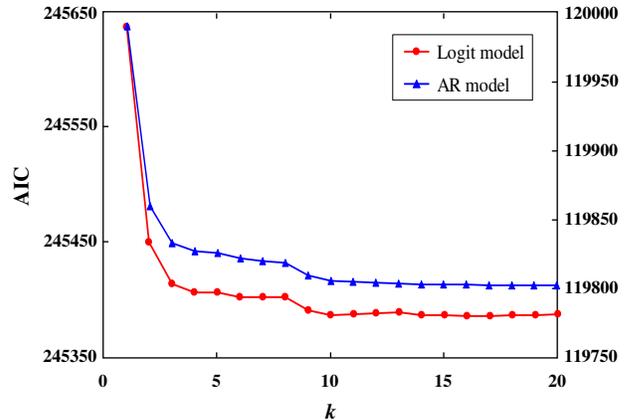}
\caption{Plot of the Akaike IC values for the AR model (the value of
the left $y$-axis) for Data $A$ and of the logit model (the value of
the right $y$-axis) for Data $B$; in each case, the Akaike IC value
decreases gradually as the order of model grows.} \label{f4}
\end{figure}

\begin{table}[]
\caption[0]{Values of the correct match ratio from the simulation
results of  Data A and Data B.}
\begin{tabular}{lcr}
\hline KOSPI  & Data A & Data B   \\ \hline
NP &$133,823$ & $86,561$ \\
\hline
AR     model        &$65.3\%$  &$52.2\%$    \\
Probit model        &$51.4\%$  &$50.2\%$    \\
Logit model         &$48.6\%$  &$49.8\%$    \\
\hline
\end{tabular}
\end{table}

For simplicity, we used the AR, logit, and probit models to analyze
the $X(t)$ series for high-frequency tick data of the Korean
financial market. As shown in Fig. $4$, we found that the Akaike IC
values for the AR and logit models decrease gradually as the order
of the models increases.Because the Akaike IC for the three models
has approximately the same value in a range larger than the order of
$k=15$, we consider this value to be the minimum value; in addition,
this value is similar to the $10$th order of the AR model of the
yen-dollar exchange rate $[11]$. Hence, the function shape of the
logit model is similar to that of the probit model, and each
probability structure tends to move continuously in the same
direction. By minimizing the Akaike IC value of our model, we were
also able to calculate the correct match ratio. Table $2$ shows the
values of the correct match ratios for Data $A$ and Data $B$. The AR
model of Data $A$ has a higher value than other models for the
correct match ratio; in contrast, the logit model of Data $B$ has a
smaller value.

%\section {Conclusions }

In conclusion, we used the AR, logit, and probit models to determine
the probability structure of high-frequency tick data of the KOSPI
in the Korean financial market. The value of our conditional
probability of the KOSPI is slightly greater than that of the
yen-dollar exchange rate. Our results show that the Korean financial
market is slightly unstable and less systematic than other financial
markets, though the results may be related to actual transactions of
all assets. In addition, by using the AR, probit, and logit models,
we deduce that the forecasted (or simulated) sign is equal to the
sign of the actual returns. This deduction enables us to obtain the
correct match ratio. Moreover, because the match ratio is always
greater than 0.5, we can conclude that our model has an improved
forecasting capability. The AR model, which is expected to have a
higher predictable value only in the Korean financial market,
robustly supports the future predictability of price movement trends
in financial markets. We also note that, with nonlinear models of
data analysis, international finance theories can offer an enhanced
interpretation of results. For the past decade, many econophysical
investigations have led to greater appreciation of, and insight
into, scale invariance and the universality of statistical
approaches to physics and economics. Our results should encourage
interdisciplinary research of physics and economics.

%This work was supported by the Korea Research Foundation Grant
%funded by the Korean Government (MOEHRD) (KRF-2005-041-C00183).

%\begin{acknowledgements}
%This work was supported by Pukyong National University Research
%Fund in 2005.
%This work was supported by Korea Research Foundation
%Grant(KRF-2004-002-B00026).
%\end{acknowledgements}

\end{document}